\documentclass{emulateapj}
 
\newcommand{\etal}{et\,al.}

\newcommand{\lsim}{\raise0.3ex\hbox{$<$}\kern-0.75em{\lower0.65ex\hbox{$\sim$}}}

\newcommand{\msun}{M$_{\odot}$}

\newcommand{\kms}{km\,s$^{-1}$}

\begin{document}
\slugcomment{Accepted for Publication in the Astrophysical Journal}

\title{On the Origin of the Supergiant HI Shell and Putative Companion
  in NGC 6822\footnote{Based on observations made with the NASA/ESA
  Hubble Space Telescope, obtained from the Data Archive at the Space
  Telescope Science Institute, which is operated by the Association of
  Universities for Research in Astronomy, Inc., under NASA contract
  NAS 5-26555}}

\author{John M. Cannon$^{1}$, 
Erin M. O'Leary$^{1}$, 
Daniel R. Weisz$^{2}$,
Evan D. Skillman$^{3}$,
Andrew E. Dolphin$^{4}$,
Frank Bigiel$^{5}$,
Andrew A. Cole$^{6}$,
W.J.G. de Blok$^{7}$,
Fabian Walter$^{8}$,}
\affil{\begin{scriptsize}
$^{1}$Department of Physics \& Astronomy, Macalester College, 1600 Grand Avenue, Saint Paul, MN 55105; jcannon@macalester.edu\\
$^{2}$Department of Astronomy, University of Washington, Seattle, WA 98195\\
$^{3}$Minnesota Institute for Astrophysics, University of Minnesota, Minneapolis, MN 55455, USA\\
$^{4}$Raytheon Company, 1151 East Hermans Road, Tucson, AZ 85706\\
$^{5}$Institut f\"ur Theoretische Astrophysik, Universit\"at Heidelberg, Albert-Ueberle-Str. 2, 69120 Heidelberg, Germany\\
$^{6}$University of Tasmania, School of Mathematics \& Physics, Private Bag 37, Hobart, Tasmania 7001\\
$^{7}$Department of Astronomy, University of Cape Town, Rondebosch 7700, South Africa\\
$^{8}$Max-Planck-Institut f{\"u}r Astronomie, K{\"o}nigstuhl 17, D-69117, Heidelberg, Germany
\end{scriptsize}}

\begin{abstract}

We present new {\it Hubble Space Telescope} Advanced Camera for
Surveys imaging of six positions spanning 5.8 kpc of the HI major axis
of the Local Group dIrr NGC 6822, including both the putative
companion galaxy and the large HI hole.  The resulting deep color
magnitude diagrams show that NGC 6822 has formed $>$50\% of its stars
in the last $\sim$5 Gyr.  The star formation histories of all six
positions are similar over the most recent 500 Myr, including
low-level star formation throughout this interval and a weak increase
in star formation rate during the most recent 50 Myr.  Stellar
feedback can create the giant HI hole, assuming that the lifetime of
the structure is longer than 500 Myr; such long-lived structures have
now been observed in multiple systems and may be the norm in galaxies
with solid-body rotation.  The old stellar populations (red giants and
red clump stars) of the putative companion are consistent with those
of the extended halo of NGC 6822; this argues against the
interpretation of this structure as a bona fide interacting companion
galaxy and against its being linked to the formation of the HI hole
via an interaction.  Since there is no evidence in the stellar
population of a companion galaxy, the most likely explanation of the
extended HI structure in NGC 6822 is a warped disk inclined to the
line of sight.

\end{abstract}						

\keywords{galaxies: evolution --- galaxies: dwarf --- galaxies:
  irregular --- galaxies: individual (NGC 6822)}

\section{Introduction}
\label{S1}

Holes and shells in the neutral interstellar medium of star-forming
galaxies have been challenging to interpret for decades.  While
alternative creation mechanisms exist, and the details of the
energetics and multi-phase interactions are very complex (see, e.g.,
discussion in {Tenorio-Tagle \& Bodenheimer 1988}\nocite{tt88}),
recent results support a scenario where energetic processes associated
with stellar evolution (feedback) are the primary mechanism by which
such structures are created (a very short and incomplete list would
include {Weisz \etal\ 2009a}\nocite{weisz09a}, {Weisz
  \etal\ 2009b}\nocite{weisz09b}, {Cannon
  \etal\ 2011a}\nocite{cannon11a}, {Cannon
  \etal\ 2011b}\nocite{cannon11b}, {Warren
  \etal\ 2011}\nocite{warren11}, and the numerous references therein).
Many of these results have been attained in low-mass dwarf galaxies,
where the lack of large-scale dynamical processes (i.e., spiral
density waves) allows the characteristic timescale of the gas
structures to extend for long periods of time (longer than a few
hundred Myr in the aforementioned studies). The stellar feedback cycle
can be considered one of the primary internal catalysts of galaxy
evolution.

A second driver of rapid galaxy evolution is gravitational
interactions with companion systems. Interactions between massive
galaxies can cause powerful starburst episodes (e.g., {Heckman
  \etal\ 1990}\nocite{heckman90}); similarly, interactions between
dwarf and massive galaxies can be dramatic (e.g., the Magellanic
Stream being pulled from the SMC/LMC system by interactions with the
Milky Way; {Putman \etal\ 1998}\nocite{putman98}).  However,
interactions between low-mass systems, which were frequent in the
early universe, are less common at low redshift.  Part of the reason
is that most present day star forming, gas-rich dwarf galaxies are found in
regions of comparative isolation or in loose associations such as the
Local Group.  Thus, ``isolated'' dwarf galaxies (i.e., not satellites
of massive systems) that are undergoing interactions with other
low-mass objects are rare, with only a few well-studied examples (e.g., VCC
848 - {Gallagher \& Hunter 1989}\nocite{gallagher89}).  Interactions
between companion low-mass systems is a relatively unexplored driver of dwarf
galaxy evolution.

At a distance of 492 $\pm$ 20 kpc (1\arcsec\ $=$ 2.4 pc;
{Lee \etal\ 1993}\nocite{lee93}; Dolphin \etal\ in preparation), NGC\,6822 is
among the nearest star forming dIrrs in the Local Group.  This system
is gas-rich, dark-matter dominated (de~Blok \& Walter 2000), and
resembles the SMC in size and luminosity (M$_{\rm B} \sim$ --15.8;
{Hodge \etal\ 1991}\nocite{hodge91}). It is not associated with the
concentrations of galaxies surrounding M\,31 or with the Milky Way
({Mateo 1998}\nocite{mateo98}).  While the system is well-studied
(more than 600 journal references to date), and is in our cosmological
backyard, the internal and external processes that drive its evolution
remain poorly constrained.

\begin{figure*}
\epsscale{0.785}
\plotone{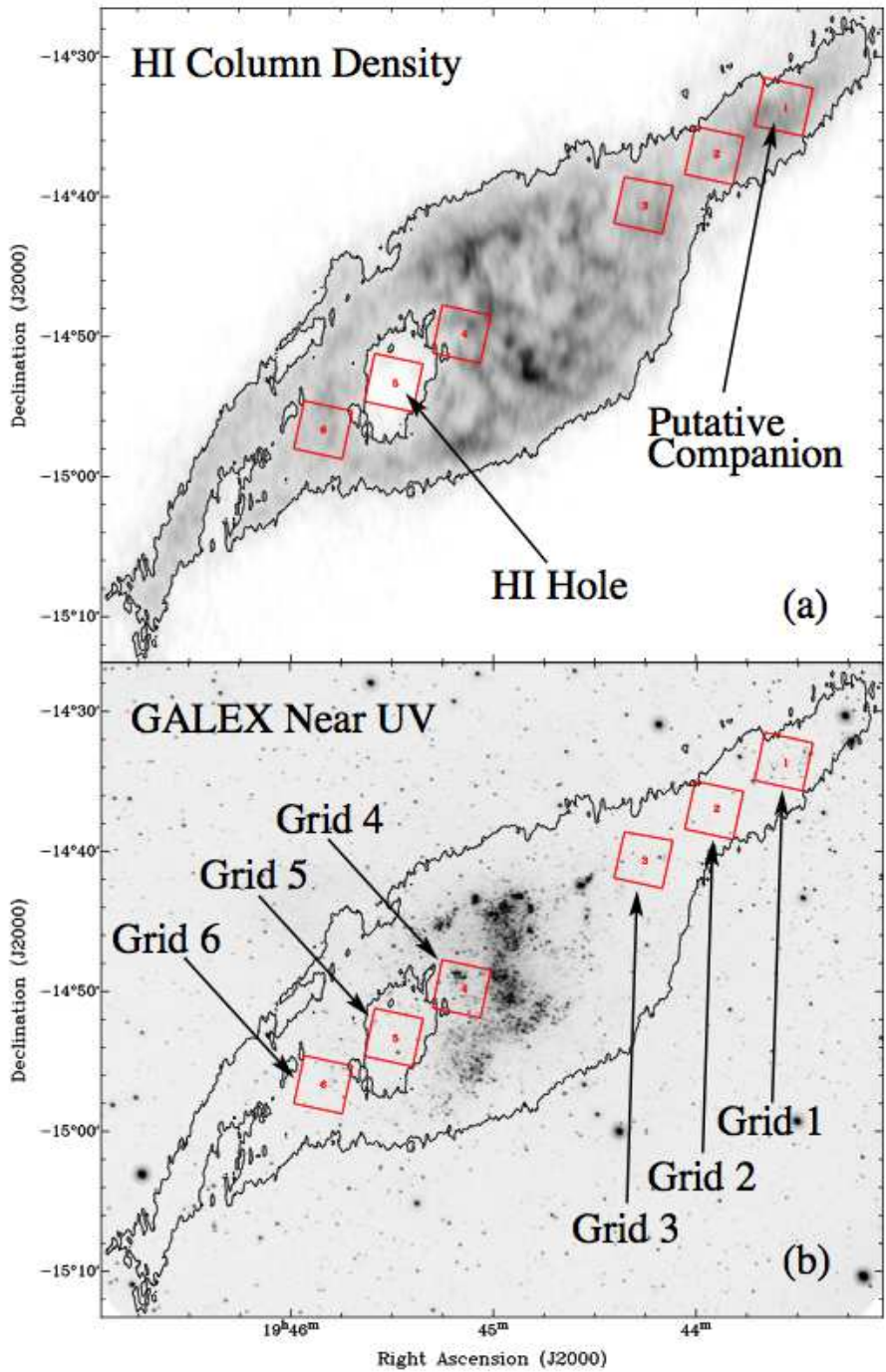}
\epsscale{1.0}
\caption{HI column density image (a; see de~Blok \& Walter 2000) and
  archival GALEX image (b) of NGC 6822, with the six observed grid
  positions overlaid in red and labeled from ``1'' (northwest) to
  ``6'' (southeast).  Grid 1 contains the putative companion and grid
  5 is located within the HI hole.  The black contour highlights the
  HI column density at the 5\,$\times$\,10$^{20}$ cm$^{-2}$ level.}
\label{figcap1}
\end{figure*}

NGC\,6822 provides an opportunity to study a system that is currently
being shaped by both internal (stellar feedback) and external
(low-mass gravitational interaction) processes.  The neutral gas disk
is dominated by an intricate structure of holes and shells (see Figure
1, and {de~Blok \& Walter 2000}\nocite{deblok00}, 2006). The most
impressive feature is a large hole spanning some 2 kpc $\times$ 1.4
kpc (labeled ``HI Hole'' in Figure~\ref{figcap1}); this is one of the
largest known holes in the ISM of a nearby dwarf galaxy.  Equally
remarkable is the presence of a putative small companion galaxy
(M$_{\rm HI}$ $\sim$ 10$^7$ \msun) that appears to be interacting with
the main disk (labeled ``Putative Companion'' in Figure 1).  de~Blok
\& Walter (2000) demonstrate a significant offset in velocity between
this northwest HI cloud and NGC\,6822, arguing for the object being a
physically separate low-mass system (see further discussion below).
NGC\,6822 is thus exceptional in that it has both a giant HI hole and
a putative low-mass companion in close proximity; it is one of the
only known galaxies with both properties.

\begin{figure*}
\epsscale{0.8}
\plotone{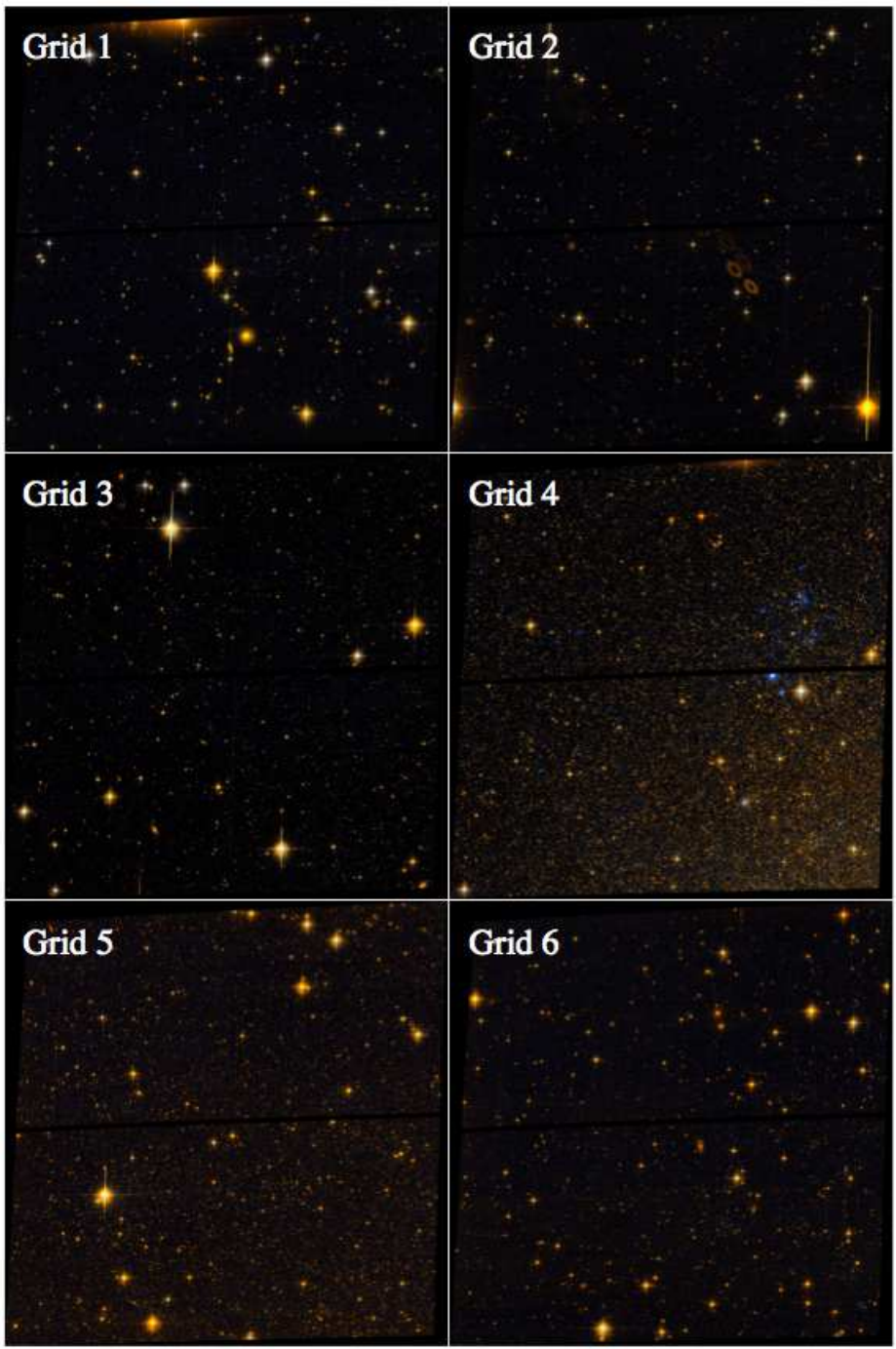}
\epsscale{1.0}
\caption{Two-color images of the six grid fields; each image is the
  size of the ACS field of view (202\arcsec\,$\times$\,202\arcsec) and
  is shown in an un-rotated orientation.  While the stellar density
  varies considerably from grid to grid, there are blue stars clearly
  detected in all locations probed by these images.}
\label{figcap2}
\end{figure*}

The ground-based optical study by \citet{deblok06} used deep {\it
  Subaru} data covering the entire HI disk to resolve the bright O and
B stars in NGC\,6822.  That study revealed the surprising presence of
numerous young stars (ages $<10^8$\,yr) throughout the entire HI disk,
well beyond $R_{25}$. Young blue stars were found associated with the
putative companion galaxy, indicating that star formation (SF) has
been ongoing there for at least the last $10^8$ yr (see {de~Blok \&
  Walter 2006}\nocite{deblok06}).  Further, there are also relatively
young stars located in the outer HI disk.

\begin{deluxetable*}{lccccccc}  
\tablecaption{{\it HST} Observations of NGC 6822} 
\tablewidth{0pt}  
\tablehead{ 
\colhead{Grid} &\colhead{RA}      &\colhead{DEC}     &\colhead{F475W t$_{\rm int}$} &\colhead{F814W t$_{\rm int}$} &\colhead{\# of Matched} &\colhead{F475W 50\% } &\colhead{F814W 50\% }\\
\colhead{} &\colhead{}      &\colhead{}     &\colhead{} &\colhead{} &\colhead{} &\colhead{Completeness} &\colhead{Completeness}\\
\colhead{\#}   &\colhead{(J2000)} &\colhead{(J2000)} &\colhead{(sec)}              &\colhead{(sec)}              &\colhead{Stars}         &\colhead{(mag)}                   &\colhead{(mag)}}
\startdata      
\vspace{0.0 cm} 
1 &19$^{\rm h}$43$^{\rm m}$33.50$^{\rm s}$  	&$-$14\degr33$^{\rm m}$41.0$^{\rm s}$  &886  &1346  &4715 &26.8 &25.5 \\
2 &19$^{\rm h}$43$^{\rm m}$54.02$^{\rm s}$  	&$-$14\degr37$^{\rm m}$09.6$^{\rm s}$  &1119 &1118  &5824 &26.8 &25.5 \\
3 &19$^{\rm h}$44$^{\rm m}$14.87$^{\rm s}$  	&$-$14\degr40$^{\rm m}$39.7$^{\rm s}$  &1119 &1118  &15850 &26.8 &25.5 \\
4 &19$^{\rm h}$45$^{\rm m}$10.25$^{\rm s}$  	&$-$14\degr49$^{\rm m}$46.4$^{\rm s}$  &1119 &1118  &195588 &26.2 &24.9 \\
5 &19$^{\rm h}$45$^{\rm m}$30.26$^{\rm s}$  	&$-$14\degr53$^{\rm m}$14.3$^{\rm s}$  &1119 &1118  &52449 &26.6 &25.3 \\
6 &19$^{\rm h}$45$^{\rm m}$51.13$^{\rm s}$  	&$-$14\degr56$^{\rm m}$36.7$^{\rm s}$  &886  &1346  &13435 &26.8 &25.4 \\
\enddata     
\label{t1}
\end{deluxetable*}   

The goal of the present investigation is to study the underlying
stellar populations in the giant HI hole and in the putative
companion, thus probing both the internal and external processes that
are driving the recent evolution of NGC 6822.  New {\it Hubble Space
  Telescope} ({\it HST}) Advanced Camera for Surveys (ACS; {Ford
  \etal\ 1998}\nocite{ford98}) imaging is acquired in six positions
along the HI major axis.  Each position is labeled in
Figure~\ref{figcap1}, from ``grid 1'' through ``grid 6'', moving from
NW to SE; note that the grid 1 position is coincident with the
putative companion, while the grid 5 position is coincident with the
HI hole.  The extracted resolved stellar photometry reaches well below
the red clump, allowing an accurate measurement of the global SFH
as well as the recent SFH (with high time resolution over the past
$\sim$500 Myr).  The SFHs we derive provide strong temporal constraints
on the creation of the various structures observed in the optical and
in HI.  This is a unique opportunity to study multiple important
evolutionary processes in dwarf galaxies.

\section{Observations and Data Reduction}
\label{S2}

Six positions spanning the HI major axis of NGC 6822 were observed for
one orbit each in Cycle 18 for program GO-12180 (P.I. Cannon); these
six positions, labeled as grid 1 through grid 6, are shown by red
boxes in Figure~\ref{figcap1}.  The {\it HST}/ACS was used to acquire
images in the F475W and F814W filters in two visits (grids 1, 2, and 3
observed on October 22, 2010; grids 4, 5, and 6 observed on March
28-29, 2011).  Two images were acquired with each filter at each
position, with integration times differing slightly between grids (886
and 1346 seconds for F475W and F814 in grids 1 and 6, and 1119 and
1118 seconds for F475W and F814 in grids 2, 3, 4 and 5).  No dithering
strategy was applied; each filter's observation was CRSPLIT into two
images to facilitate the removal of cosmic rays.  Hot pixels are
easily differentiated from stars using our photometric measurement
tools (see below).  Figure~\ref{figcap2} shows color representations
of each {\it HST} field of view; the presence of both young blue stars
and a diffuse red stellar population is obvious in all grid positions.

\begin{figure*}
\epsscale{1.0}
\plotone{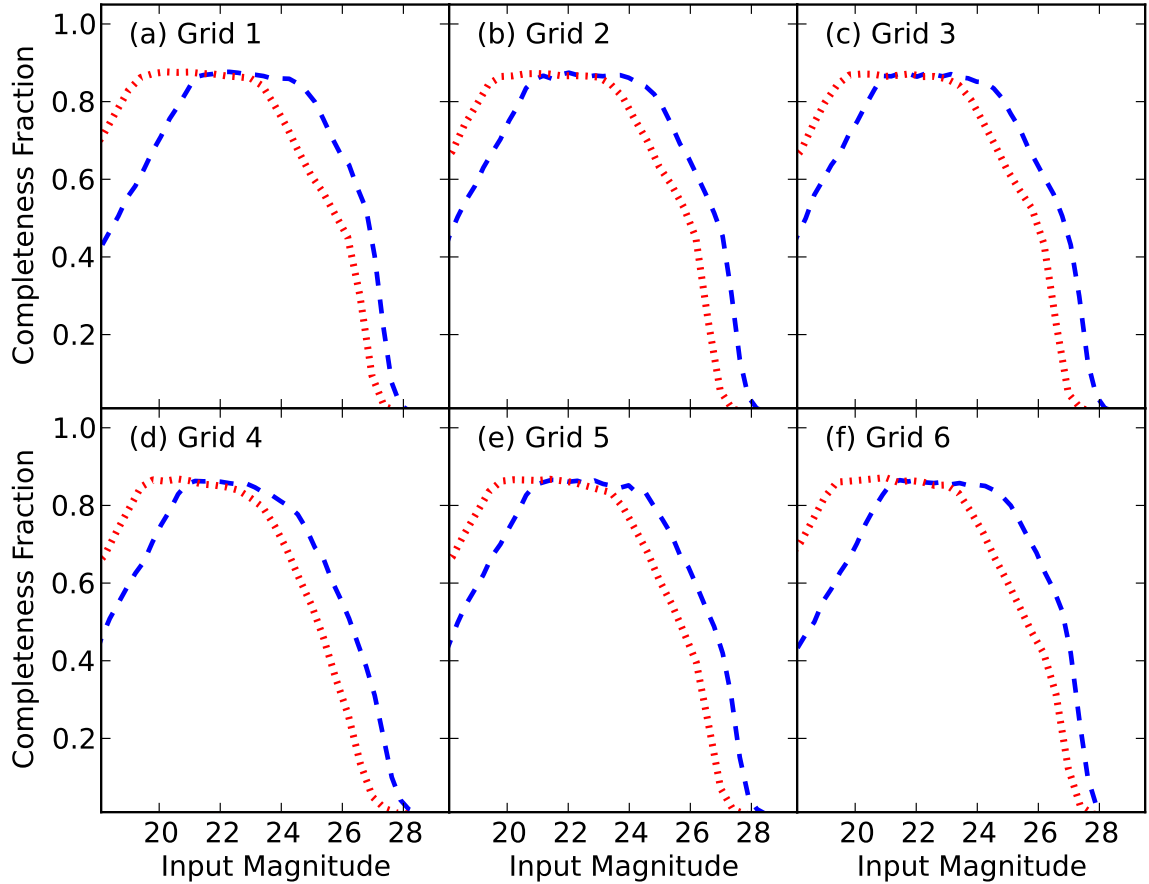}
\epsscale{1.0}
\caption{Photometric completeness as a function of magnitude for each
  of the six grid positions.  The blue and red lines represent the
  functions for the F475W and F814W filters, respectively.  The
  photometric depths and completeness functions are nearly identical
  for all six grid positions.}
\label{figcap3}
\end{figure*}

We performed stellar PSF photometry using the DOLPHOT software
package, a modified version of HSTPHOT \citep{dolphin00} with an ACS
specific module. To separate well-measured stellar sources from
artifacts, we applied quality criteria to the raw photometric catalogs
in each field. Specifically, a well-measured star satisfied the
following criteria: SNR$_{\rm F475W}$ $>$ 5.0 and SNR$_{\rm F814W}$
$>$ 5.0, (sharp$_{\rm F475W}$ $+$ sharp$_{\rm 814}$)$^2$ $<$ 0.075,
and (crowd$_{\rm F475W}$ $+$ crowd$_{\rm F814W}$) $<$ 1.0.  The 50\%
completeness levels and numbers of matched stars in each grid location
are given in Table~\ref{t1}.  To characterize the observational
uncertainties and completeness functions, in each field we performed
500,000 artificial star tests. In order to compute the completeness
function, the artificial star tests were culled using identical
criteria to the photometry. The results of these tests are shown in
Figure~\ref{figcap3}, which plots the photometric completeness as a
function of magnitude for each of the six grid positions.  The
photometric depths and completeness functions are nearly identical for
all six grid positions.

\begin{figure*}
\epsscale{1.0}
\plotone{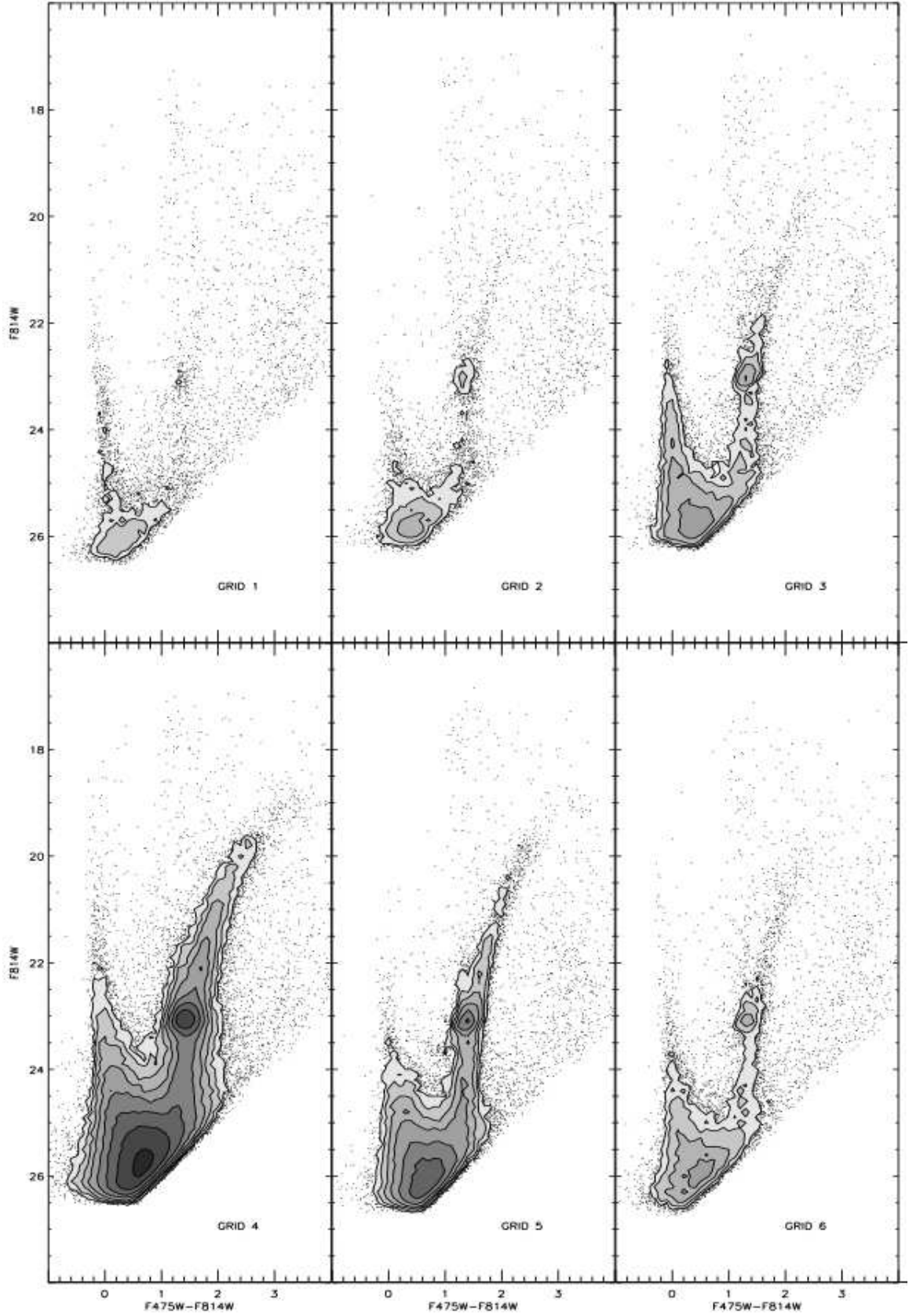}
\epsscale{1.0}
\caption{Hess diagram CMDs of each grid, as labeled; each grid is
  corrected for MW foreground extinction via the best-fit value from
  SFH analysis.  Contours are at the same levels (10, 20, 40, 80, 160,
  320, 640, 1280, 2560 stars decimag$^{-2}$) in all panels.}.
\label{figcap4}
\end{figure*}

Figure~\ref{figcap4} presents the color magnitude diagram (CMD) of
each grid position using a density contour approach (i.e., a ``Hess
diagram'').  Our photometry is complete well below the red clump
and fully samples the blue helium burning (BHeB) region of the CMD in
each grid (see {Weisz \etal\ 2008}\nocite{weisz08} for a complete
discussion of these various CMD regions).  Note the differences in
stellar density from one grid to the next (e.g., grid 4 contains
$\sim$40 times more matched stars than grid 1).

An accurate CMD of a galaxy's stellar populations allows the
derivation of the intensity of SF as a function of both time and of
location, i.e., a star formation history (SFH). To derive the SFH, we
apply a maximum-likelihood technique thatits the observed features
of the CMD with a set of synthetic CMDs. The SFH of the best fitting
synthetic CMD is the most probable SFH of the observed stellar
population.  We refer the reader to \citet{dolphin02},
\citet{dolphin03}, and Dolphin (2012, in preparation) for detailed
discussion and demonstrations of this technique and its ability to
differentiate flat SFHs from those with temporal features.

The synthetic CMDs are a linear combination of simple stellar
populations that have been convolved with observational biases as
determined by artificial star tests. We generate each synthetic
population by assuming a stellar initial mass function and a binary
star fraction, and designating a number and resolution of bins in both
age and metallicity. Further, the observed and synthetic CMDs are
binned to facilitate statistical comparisons between the two.

In the case of NGC 6822, we have adopted a Kroupa IMF ({Kroupa
  2001}\nocite{kroupa01}) within a mass range of 0.15 to 120 \msun, a
binary fraction of 0.35, and the Padova stellar evolution models
({Girardi \etal\ 2002}\nocite{girardi02} and {Girardi
  \etal\ 2010}\nocite{girardi10})).  To balance computational
efficiency with the desire for a detailed SFH, we selected 36 time
bins with uniform logarithmic spacing between $\log(t) =$ 6.6 and
10.15 and an isochronal metallicity range from 0.0001 Z$_{\odot}$ to
0.0019 Z$_{\odot}$ with a resolution of 0.1 dex. The data do not
contain significant main sequence turn off features for populations
older than $\sim$5 Gyr, and we thus have only poor constraints on the
lifetime chemical evolution.  Therefore, we required that the
metallicities monotonically increase to minimize potential
non-physical chemical models resulting from the RGB age-metallicity
degeneracies \citep[e.g.,][]{gallart05}. We have designated the 50\%
completeness limits as the faintest photometric depths considered.

Intervening Milky Way foreground stars can contaminate the observed
CMD.  To mitigate this effect, we have included a model foreground CMD
in the SFH measurements.  This CMD was constructed with the Dartmouth
isochrones \citep{dotter08}, which extend to slightly lower masses of
0.1 \msun, and follow the color and magnitude distributions detailed
in \citet{dejong10}.  Finally, given the relative sparseness of the
TRGB populations in each field, we fixed the distance to previous TRGB
determinations (see \S~\ref{S1}). However, we allowed the code to
solve for the optimal extinction value to each field, as NGC~6822
appears to have significant variations in extinction from field to
field.

In this study, we are interested in comparing the SFHs derived for
each grid position; each of the six CMDs contains data of comparable
photometric depths and similar error functions (see
Figure~\ref{figcap3}).  The similarity of data quality at each
position means that the uncertainties on the SFHs reflect the
observational errors and the random uncertainties due to the number of
stars.  The random uncertainties are calculated by performing 50 Monte
Carlo tests that use a Poisson random noise generator to create a
random sampling of the best-fit CMD.  We do not consider the
systematic uncertainties introduced by choice of stellar evolution
models that are typically calculated in SFH derivations; in the limit
of comparable quality data these choices do not affect the
differential measurements from one position to another
\citep[e.g.,][]{weisz11}.

\section{Star Formation Histories in the Six Grid Positions}
\label{S3}

\begin{figure*}
\epsscale{0.98}
\plotone{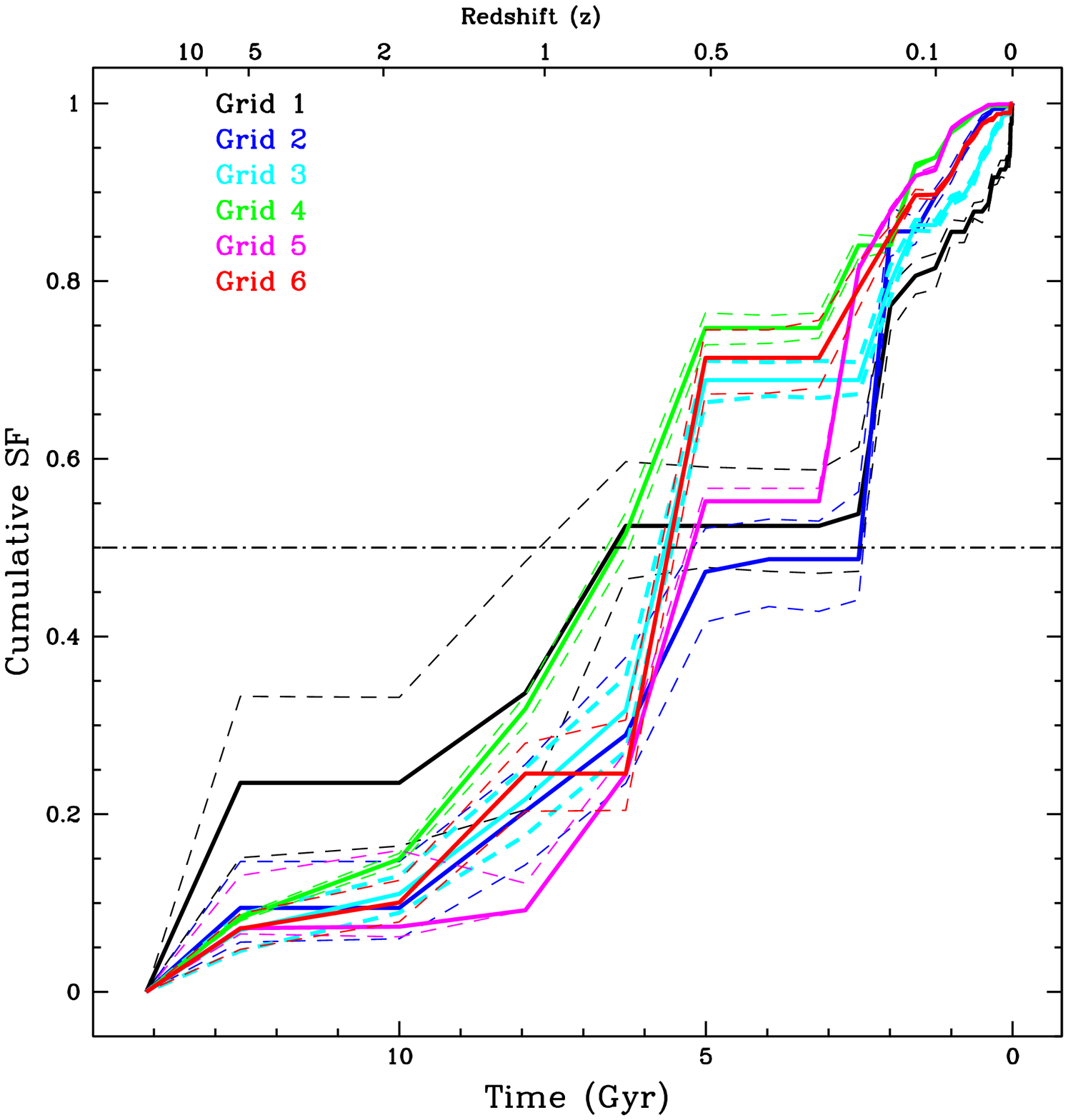}
\epsscale{1.0}
\caption{The cumulative SFHs of each grid.  The solid lines represent
  the best-fit cumulative SFH for each grid; the dotted lines
  represent the 1\,$\sigma$ error bounds (technically the bounds at
  the 16$^{\rm th}$ and 84$^{\rm th}$ percentiles) derived from Monte
  Carlo simulations.  When a given region reaches a cumulative SF of
  0.5, then 50\% of the total stellar mass in that region has
  formed. These plots demonstrate that NGC 6822 has formed $\sim$50\%
  of its stellar mass within the most recent 5 Gyr.  The redshifts at
  the top of the plot were computed assuming a WMAP7 cosmology
  \citep{jarosik11}.}
\label{figcap5}
\end{figure*}

The cumulative SFHs of all six grid positions are presented in
Figure~\ref{figcap5}. This shows, for lookback times beginning at the
Big Bang and decreasing until the present day, the fraction of the
total stellar mass in each field that formed at or before a given
time. All grid positions are normalized to a cumulative value of unity
at the present day.  When a given region reaches a cumulative SF of
0.5, then 50\% of the total stellar mass in that region has formed; we
hereafter refer to this timescale, measured back from the present day,
as $\tau_{\rm 50}$.  The solid lines represent the best-fit cumulative
SFH for each grid; the dotted lines represent the 1\,$\sigma$ error
bounds (technically the bounds at the 16$^{\rm th}$ and 84$^{\rm th}$
percentiles) derived from Monte Carlo simulations.

\begin{figure*}
\epsscale{1.0}
\plotone{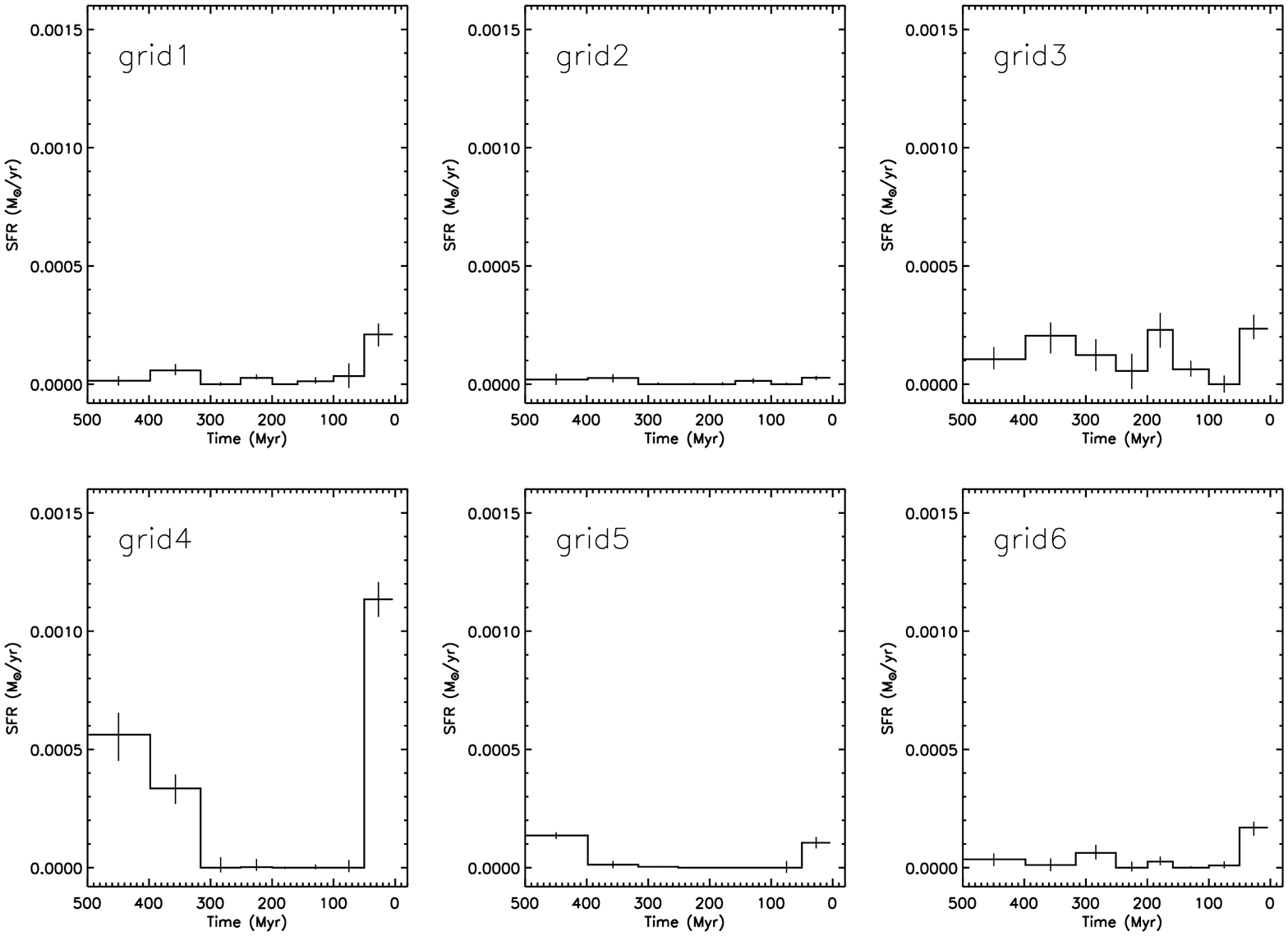}
\epsscale{1.0}
\caption{Recent SFHs in each grid over the most recent 500 Myr; note
  that time increases from 0 Myr (present day) to 500 Myr leftward on
  the abscissae. The general shapes of the SFHs are similar in all
  grids (i.e., the periods of relative activity are contemporaneous),
  although the strength of SF varies from one grid position to the
  next.}
\label{figcap6}
\end{figure*}

While some variations exist, the cumulative SFHs of the six grid
positions are qualitatively consistent with one another from the
Hubble time up until $\sim$6 Gyr ago. This is strong evidence for the
entire NGC 6822 system evolving as a single entity during this period.
If the putative companion in grid 1 were in fact undergoing its first
episode of SF at the present time, then such a signature would be very
obvious in the plot as a long period of quiescence followed by a rapid
increase in the cumulative SF in recent epochs.  No such pattern is
seen for the stars detected in grid 1.  We note that our age
resolution is most coarse for these ancient bins; however, this caveat
strengthens our interpretation.  Even if the details of the ancient
SFH are less certain than those at recent times, Figure~\ref{figcap5} 
shows that the field containing the putative companion has formed the 
highest fraction of its stars at old ages compared to the other fields.
In fact, this is exactly what is expected if the outer field contains just 
the extended disk and not a companion.  Numerous studies of the resolved
stars of nearby dwarf galaxies have shown that the outer regions have
higher fractions of older stars \citep[e.g.,][]{dp97, dp98, jsg98, et98}.
This nearly universal characteristic of dwarf galaxies has been 
interpreted as the result of the active star-forming region of the galaxy 
shrinking with time \citep{hma03}, and has been supported by very detailed
spatially resolved SFHs \citep[e.g.,][]{hadg09}.

As Figure~\ref{figcap5} shows, NGC 6822 has formed roughly half of its
stars within the most recent 5 Gyr (i.e., $\tau_{\rm 50}$ $\simeq$ 5
Gyr).  This is broadly consistent with the value $\tau_{50} = 6.9$~Gyr
reported by \citet{orban08} in a reanalysis of archival {\it HST}
images near the center of NGC~6822, who found that 57\% of the
lifetime star formation took place in the past 5~Gyr.  Our derived
$\tau_{\rm 50}$ value can be compared with measurements of the same
timescale for other nearby galaxies. Using a sample of seven Local
Group dIrr galaxies, \citet{weisz11} calculates a characteristic
timescale of $\sim$6 Gyr for the formation of 50\% of the stellar
mass; NGC 6822 thus is very similar to other Local Group dIrrs.
Interestingly, the $\tau_{\rm 50}$ level for 25 other dIrr galaxies
from the \citet{weisz11} sample (mostly outside the Local Group) show
an average $\tau_{\rm 50}$ that is about a factor of two longer (i.e.,
a greater fraction of early star formation).

\begin{figure*}
\epsscale{1.0}
\plotone{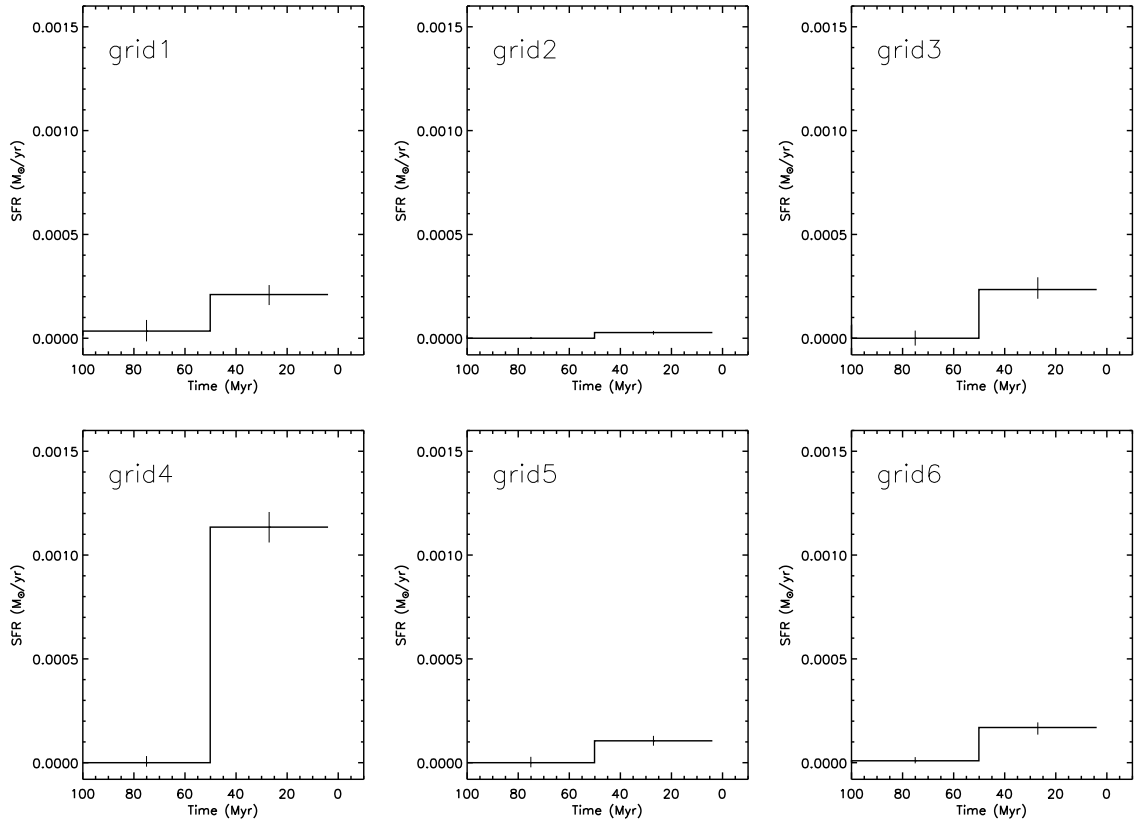}
\epsscale{1.0}
\caption{Recent SFHs in each grid over the most recent 100 Myr; note
  that time increases from 0 Myr (present day) to 100 Myr leftward on
  the abscissae. }
\label{figcap7}
\end{figure*}

The more recent cumulative SFHs (from $\sim$6 Gyr to the present) do
show some appreciable differences between grids 1, 2, and 5 and the
other three.  Specifically, grids 1, 2, and 5 show a period of
relative quiescence (i.e., little to no increase in the cumulative SF)
between $\sim$6 and $\sim$3 Gyr, while grids 3, 4, and 6 show an
increasing cumulative SF during the same interval.  Around $\sim$2-3
Gyr ago, grids 1, 2, and 5 show a significant increase in cumulative
SF.  This could be interpreted as weak evidence for an interaction
involving the material now located in grids 1 and 2 with the material
currently located in grid 5.  More specifically, this could be
interpreted as weak evidence for an interaction scenario leading to
the creation of the large HI hole at the position of grid 5.  We
discuss this hypothesis in more detail below.

The overall similarity of the cumulative SFHs of the six NGC 6822
grids, especially in the epochs more ancient than $\sim$6 Gyr, can be
interpreted as the lack of evidence for inside-out growth of its disk.
Within the errors, the outermost field is as ancient as the inner
fields.  This contrasts with the results of \citet{williams09}, who
detect the signature of increasing disk scale length with cosmic time
in M\,33.  The cumulative SFH of NGC 6822 can be considered to be
intermediate between the extreme Local Group dwarfs LGS-3 (which
formed $\sim$90\% of its stars in an intense SF event more than 11 Gyr
ago; {Hidalgo \etal\ 2011}\nocite{hidalgo11}) and Leo A (which formed
90\% of its stars more recently than 8 Gyr ago, with a peak in the
most recent$\sim$1.5-4 Gyr period; {Cole \etal\ 2007}\nocite{cole07}).

The present study focuses on the recent patterns of star formation
throughout NGC 6822. The recent SFHs (in M$_{\odot}$\,yr$^{-1}$) in
each grid location as a function of time, over the last 500 Myr and
100 Myr, are presented in Figures~\ref{figcap6} and \ref{figcap7},
respectively.  Note that since each panel presents the SFH derived
from all stars in the entire ACS field of view at each grid location,
the SFHs can also be considered in units of star formation rate (SFR)
per unit time per unit area (the
$\sim$202\arcsec$\times$202\arcsec\ field of view of the ACS
translates to $\sim$482\,$\times$\,$\sim$482 pc or $\sim$0.23
kpc$^{2}$ at the distance of 492 pc).  From the longer-duration SFHs
(Figure~\ref{figcap6}) it is clear that the SFR in each grid (with the
exception of grid 4) has been essentially flat over the last 500 Myr;
no major SF events are apparent in any grid location (except perhaps
grid 4) until the most recent $\sim$50 Myr interval.

As Figure~\ref{figcap7} shows, all six fields show evidence of an
increased rate of star formation in the last 50 Myr (although with
considerable differences in the amplitude of increase from one grid
location to the next).  Such a global increase in star formation,
which is happening on a time scale which is short compared to the
dynamical time scale ($\sim$ 140 Myr; {McQuinn
  \etal\ 2010}\nocite{mcquinn10}) is often associated with an
interaction.  Thus, we next investigate the possibility that the NW HI
concentration is a companion galaxy.

\section{On the Nature of the Putative Companion}
\label{S4}

The original interpretation of the NW region as a companion galaxy by
de~Blok \& Walter ({2000}\nocite{deblok00}, {2003}\nocite{deblok03},
{2006}\nocite{deblok06}) was based on the observed neutral gas
velocity offset (compared to the rest of the HI disk) in this region
and on the presence of a significant number of spatially coincident
young stars.  \citet{deblok06} derive a total luminosity of M$_{\rm
  B}$ $=$ $-$7.8 for the stars in the NW cloud, and estimate a total
stellar mass of $\sim$10$^5$ \msun.  This putative companion thus is
very gas-rich (M$_{\rm HI}$/L$_{\rm B}$ $>$10 depending on model
assumptions).

The similarity of the ancient SFHs in all six grid positions discussed
above appears to contradict the interpretation of this system as a
companion galaxy.  However, the more recent cumulative SF rates in
each grid position show some weak characteristics that might be
consistent with an interaction scenario.  We now use the information
on the stellar populations in the six grid positions to address the
question: is the NW region a bona fide companion galaxy that is
interacting with NGC 6822 at the present time?

\begin{figure*}
\epsscale{1.0}
\plotone{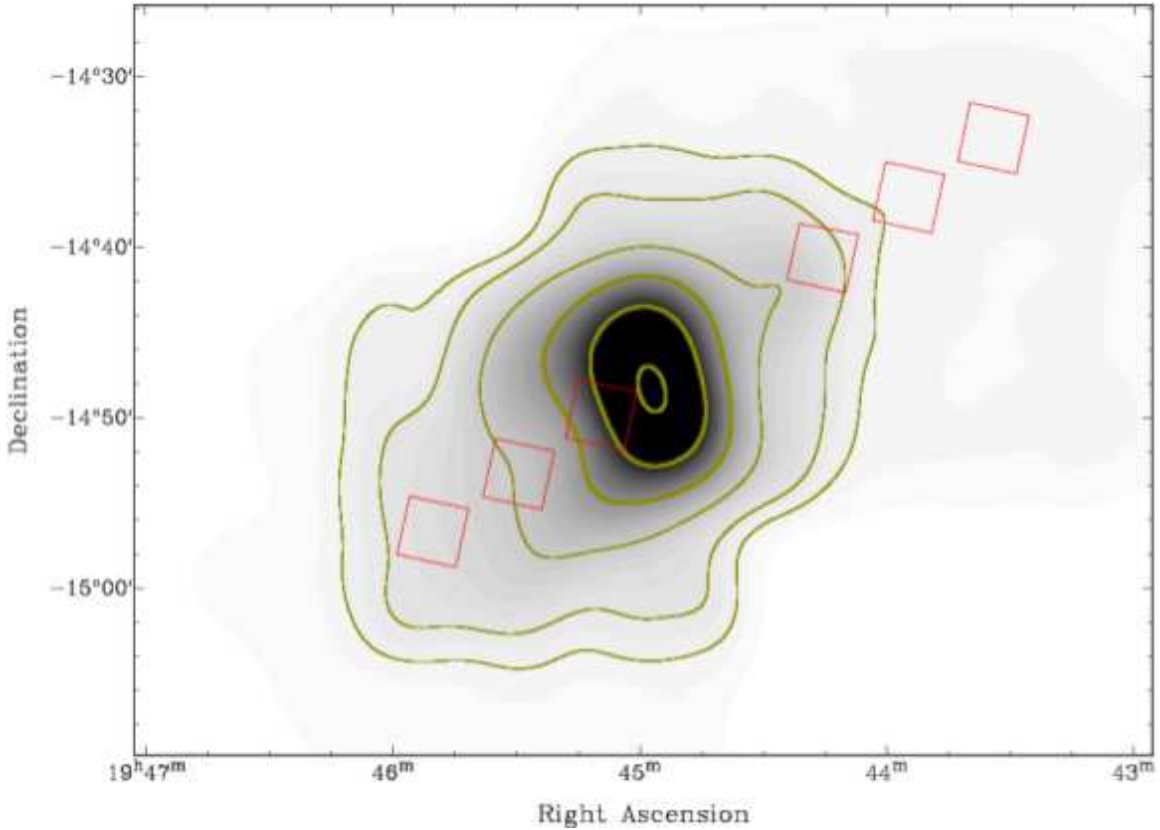}
\epsscale{1.0}
\caption{Smoothed distribution of red giant branch stars detected in
  the combined {\it Subaru} dataset from \citet{deblok06}. The six
  grid positions observed by {\it HST} are overlaid.  In grid 1 we see
  no spatial concentration of red giant branch stars indicative of the
  companion.}
\label{figcap8}
\end{figure*}

Fundamental to our interpretation is the assumption that if this is a
bona fide companion galaxy, then it should harbor an ancient stellar
population.  This assumption is based on the detection of such a
population (i.e., red clump and red giant branch stars) in every
galaxy for which adequate data exist, even extending to the most
metal-poor (e.g., I\,Zw\,18; see {{\"O}stlin 2000}\nocite{ostlin00},
{Fiorentino \etal\ 2010}\nocite{fiorentino10}, and the various
references therein) or the most extreme SFH cases (e.g., Leo\,A, with
the bulk of its SF delayed until the most recent few Gyr; see
discussion above and in {Cole \etal\ 2007}\nocite{cole07}).

If the putative companion of NGC 6822 is a bona fide dwarf galaxy,
then our ACS images of grid 1 should reveal an over-density of old
stars at this location.  Further, the ground-based {\it Subaru}
imaging of the outer regions of NGC 6822 by \citet{deblok06} should
verify this signature.  We thus explore both datasets in an attempt to
identify this ancient stellar over-density.

Figure~\ref{figcap8} shows the spatial distribution of all red giant
branch stars detected in the {\it Subaru} images; the locations of the
six ACS grids are overlaid.  While we are cognizant that crowding will
be more severe in ground-based images compared to our new ACS images,
the quality of the CMDs obtained by \citet{deblok06} allow an
unambiguous separation of the red giant branch component from the blue
plume.  This figure shows no evidence for a spatial concentration of
red giant branch stars at the location of the putative companion.

\begin{figure*}
\epsscale{1.0}
\plotone{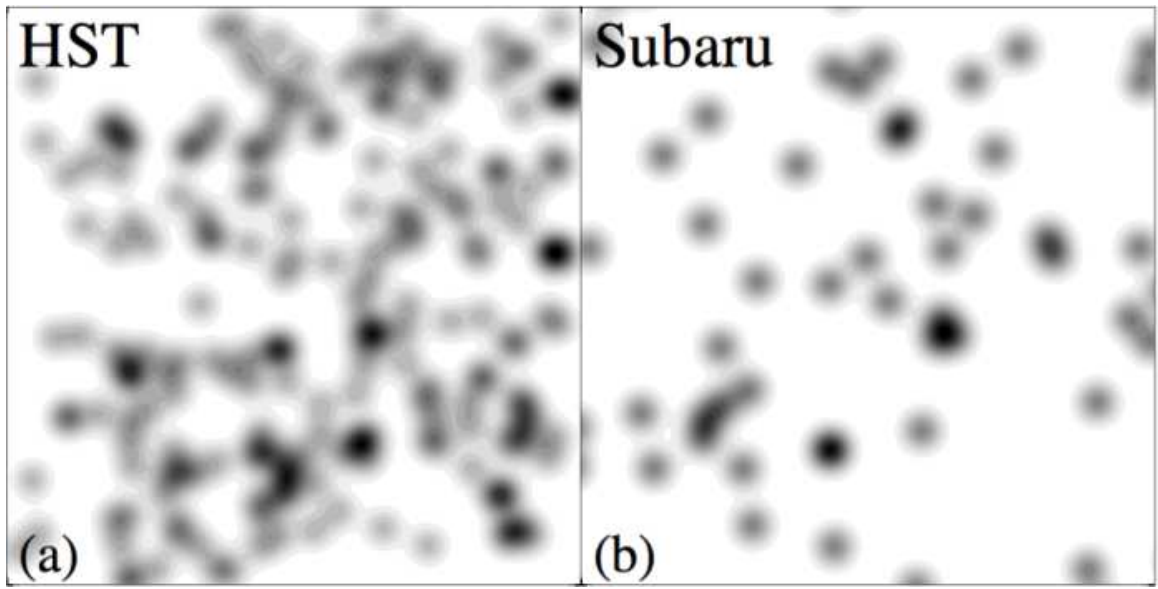}
\epsscale{1.0}
\caption{Smoothed distribution of red giant branch stars detected by
  both {\it HST} and by the combined {\it Subaru} dataset from
  \citet{deblok06}. In grid 1 we see no spatial concentration of red
  giant branch stars indicative of the companion.}
\label{figcap9}
\end{figure*}

A direct comparison of the ACS and {\it Subaru} images is challenging
not only due to the differences in crowding but also due to the
difference in photometric depth; {\it HST} will more easily detect a
faint over-density of red stars.  With these caveats in mind,
Figure~\ref{figcap9} shows the smoothed distributions of red giant
branch stars detected by both {\it HST} and {\it Subaru} in the
position of grid 1.  We stress that a one-to-one agreement between
these panels is not expected due to differences in crowding and depth.
Again, the grid 1 position shows no obvious over-density of red giant
branch stars indicative of a stellar companion.

\begin{figure*}
\epsscale{1.0}
\plotone{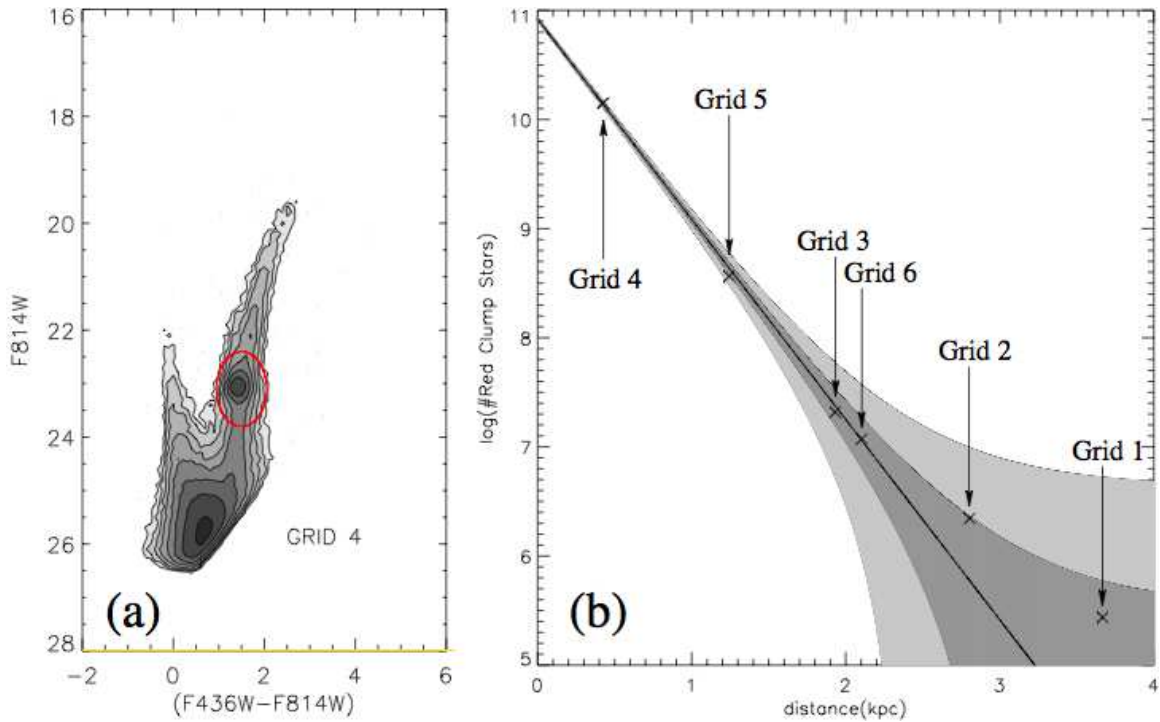}
\epsscale{1.0}
\caption{The number of red clump stars versus galactocentric radius.
  Panel (a) shows the grid 4 Hess diagram CMD (same as shown in
  Figure~\ref{figcap4}), with the red clump selection region overlaid
  in red; this same region is selected in each of the six ACS
  grids. Panel (b) shows a logartihmic plot of the number of red clump
  stars in each grid versus the projected distance from the dynamical
  center.  The solid line is an exponential fit to the data points;
  the dark and light gray regions in (b) show the 1\,$\sigma$ and
  2\,$\sigma$ bounds on this fit.  In creating this plot we have
  adopted a dynamical center position of ($\alpha$,$\delta$) $=$ (
  19$^{\rm h}$44$^{\rm m}$58.04$^{\rm s}$, $-$14\degr49$^{\rm
    m}$18.9$^{\rm s}$) (J2000) from \citet{weldrake03} and a distance
  of 492 kpc for the galaxy.  We find no evidence for an over-density
  of ancient stars at the location of grid 1.}
\label{figcap10}
\end{figure*}

It is conceivable that the ancient stellar population of the putative
companion has a sufficiently low surface brightness that it is below
the sensitivity of the {\it Subaru} data to detect. Further, it is
conceivable that the ancient stars of this putative companion are
distributed over an area larger than the projected size of the ACS
field of view, making it difficult to identify by eye.  So, as a final
test, we examine the total number of red clump stars in each of the
ACS grids.  Figure~\ref{figcap10}(a) shows the grid 4 Hess CMD with
the red clump selection region overlaid; the same selection region is
applied to the CMD and photometry of each grid.  Recall that our
photometry is complete to more than 2 magnitudes below the red clump;
such a population is easily detectable in our data.

In Figure~\ref{figcap10}(b) we plot the number of red clump stars in
each grid position as a function of distance from the dynamical center
of NGC 6822 [19$^{\rm h}$44$^{\rm m}$58.04$^{\rm s}$,
  $-$14\degr49$^{\rm m}$18.9$^{\rm s}$ (J2000) from
  \citet{weldrake03}].  The number of red clump stars per unit area
falls exponentially as a function of radius, with a scale length of
545 pc; this compares favorably with the scale length of 430 pc found
by \citet{letarte02} using the carbon star population.  The number
density of red clump stars at the grid 1 position is well within the
1\,$\sigma$ error bar of the exponential fit.  This plot is exactly
what is expected for an exponentially declining surface brightness
distribution as a function of radius.  We therefore fail to detect
evidence of an ancient stellar population that can be attributed to a
bona fide low-mass companion object at this location.  This is in
agreement with the observation by \citet{weldrake03} that the velocity
field in NGC 6822 is very highly symmetric in that the rotation curves
derived from the approaching and receding sides are nearly identical.
A true interaction would be expected to distort the velocity field in
an observable manner.  Given the high degree of symmetry, the most
likely explanation for the HI morphology is a warp in which the outer
HI is more highly inclined than the inner HI (in agreement with the
tilted ring model results found by {Weldrake et
  al. 2003}\nocite{weldrake03}).

Taken as a whole, the evidence above challenges the hypothesis that
the putative companion is a bona fide dwarf galaxy that harbors an
ancient stellar population. The origin of the global SF event during
the most recent $\sim$50 Myr (which varies in intensity from one grid
position to another) may be caused by a gravitational interaction, but
we do not find evidence that the object in the NW is the interacting
partner.  This event may simply be stochastic - a result of the small
number of young blue stars required to increase the SFR by the
observed amounts in the grids with weaker SFRs.

A powerful diagnostic of the nature of this putative companion would
be an observed metallicity difference between it and the main stellar
body.  The coarse metallicities of each field derived by our SFH show
no evidence for such an offset.  This further weakens the companion
hypothesis.  Of course, direct abundance estimators would be
preferable; in this regard, deep long-slit spectroscopy of the faint
HII regions in the putative companion and/or of the red giant branch
stars therein would be most useful in identifying any abundance
difference between this object and the main stellar body.

\section{On the Origin of the Supergiant HI Hole}
\label{S5}
\subsection{A Stellar Feedback Origin}
\label{S5.1}

The large hole in the neutral gas distribution of NGC 6822 has been
enigmatic since its discovery
\citep{gottesman77,skillman92conf,brandenburg98,deblok00}.  The size
of the structure ($\sim$2 kpc $\times$ 1.4 kpc), the lack of observed
expansion, and the offset location with respect to the high surface
brightness stellar distribution (see Figure~\ref{figcap1}) make it a
particularly important test case for the hypothesis that stellar
evolution can create such structures via coherent feedback.  We now
examine the recent SF within this structure with the particular goal
of discerning if stellar feedback processes are capable of producing
it.

The recent SFH of the HI hole field (grid 5; see Figures~\ref{figcap6}
and \ref{figcap7}) is characterized by activity in the $\sim$500-400
Myr and the most recent $\sim$50 Myr periods, with a period of
relative inactivity between.  Note that all of the SFRs are
comparatively low; even at the height of activity, the SFR only reaches
a level of order 10$^{-4}$ \msun\,yr$^{-1}$.  Since the HI structure
has no observed expansion velocity (and thus no empirical constraint
on its age), we apply the methodology developed in \citet{weisz09a} to
derive the amount of stellar feedback energy input into this region
over the most recent 500 or 100 Myr periods.  These calculations yield
9.8$\times$\,10$^{52}$ erg and 1.2$\times$\,10$^{52}$ erg for the most
recent 500 Myr and 100 Myr intervals, respectively. Note that we make
no correction for the contributions of Type Ia SNe to the feedback
energy budget, as has been shown to be important \citep{recchi06}.
These events occur over a range of timescales \citep{matteucci06} and
may occur in regions that are already partially evacuated.  Thus, the
feedback energies derived from stellar evolution should be considered
lower limits.

The ability of this feedback energy to evacuate the surrounding
material depends critically on how much energy is converted into gas
motion.  If one assumes a 100\% conversion efficiency of feedback
energy into the motion of the surrounding gas, then the feedback
energy budget over the last 500 Myr reaches $\sim$10$^{53}$ erg, the
value derived by \citet{deblok00} using an adopted age of 100 Myr and
the single-blast SN prescriptions in \citet{chevalier74}.  

Note that the required blast energies estimated in this way are highly
model dependent and thus highly uncertain \citep{warren11}.
Additionally, we stress that assuming an efficiency of 100\% is likely
not physical and is much higher than most derived values found in
nearby galaxies \citep[e.g.,][]{weisz09b,cannon11b,warren11}, as well
as higher than predictions from numerical simulations
\citep[e.g.,][]{theis92,cole94,padoan97,thornton98}.  With a more
realistic assumption of 20\% efficiency, the timescale for the
requisite feedback energy extends well past 500 Myr.

Despite the uncertainties, it is clear that the energy produced by
stars in the last 100 Myr is unlikely to have been able to produce the
HI hole.  It is likely that the energy from star formation over at
least the last 500 Myr is required to have produced the present HI
hole.  Assuming a stellar feedback origin, this implies that holes can
persist for much longer than the dynamical timescale ($\sim$140 Myr;
{McQuinn \etal\ 2010}\nocite{mcquinn10}).  HI holes with dynamical
ages of a few hundred Myr have been observed before \citep[see,
  e.g.,][]{weisz09b,cannon11b,warren11}, and have theoretical support
from simulations \citep[e.g.,][]{recchi06}.  Such long-lived
structures might be considered to be more likely in the lower surface
density outer HI disk of a system like NGC 6822 which is undergoing
solid-body rotation, since there is no rotational shear to remove the
structure.  Indeed, the models of \citet{maclow99} show that the
position with respect to the galactic midplane is an important
parameter in the evolution of expanding structures that result from
SF.  We postulate that the typical HI hole in a low-mass galaxy is in
fact long-lived, explaining, in part, the lack of observed expansion
velocities for many of the holes and shells in nearby dwarfs
\citep[e.g.][]{bagetakos11}.

\subsection{Alternative Creation Mechanisms}
\label{S5.2}

Could the HI hole have been caused by a gravitational interaction with
the putative companion?  In such a scenario, we would expect a global
burst of SF in both the hole (grid 5) and the companion (grid 1) on a
timescale compatible with the dynamical time.  While we do see a weak
SF signature within the most recent $\sim$50 Myr (although the
relative increases in SFRs are small in both grid positions), the
current relative separation of the HI hole and the putative companion
($\sim$34\arcmin) would require a relative velocity of $\sim$95
km\,s$^{-1}$ over this 50 Myr timescale in order for the events to be
causally connected.  Assuming that neutral gas was involved in this
interaction, such a signature would be very obvious in the HI data and
is inconsistent with the observed velocity structure presented in
\citet{deblok00} and \citet{weldrake03}.  When coupled with the lack
of evidence for an ancient stellar population at the location of grid
1 (see \S~\ref{S4}), we conclude that an interaction with the putative
companion is an unlikely creation mechanism for the giant HI hole.

The interaction scenario remains promising, however, based on the
morphology of the neutral gas alone.  The extensions of the HI disk to
the NW and SE could be indicative of tidal features that might arise
during an interaction.  The involvement of any other Local Group
galaxy in such a potential interaction seems unlikely due to the
galaxy's relative isolation.  \citet{deblok00} unsuccessfully searched
a 10 square degree area around NGC 6822 for companions detected by
HIPASS \citep{meyer04}.  However, we note that NGC 6822 has a negative
heliocentric velocity and thus is mildly confused with Galactic HI
emission.  Subsequent observations with the Parkes 64\,m multi-beam
receiver \citep{weldrake03} show that most of the Galactic emission
can be separated from that within NGC 6822, although a small velocity
range ($\sim$10 \kms) centered in the SE region could contain moderate
Galactic contamination.  A small perturbing HI cloud thus remains a
viable, though unlikely, mechanism for the creation of the HI hole and
the extended tidal structure of NGC 6822.

Various other creation mechanisms of the HI hole could be considered,
although none is as successful in explaining the current properties of
NGC 6822 as a stellar feedback origin.  For example, ram pressure
could preferentially remove the neutral gas from a low-density outer
disk region \citep[e.g.,][]{bureau02}.  While the size of the giant HI
hole, and the lack of similar-scale structures elsewhere in the outer
disk argue against ram pressure stripping as the sole creation
mechanism for the giant HI hole, \citet{bureau02} argue that ram
pressure has the capacity to enlarge pre-existing holes and lower
their creation energies.  This would help to ``bridge the gap''
between the observed star formation rate and that required to create
the holes, thus resulting in lower creation ages.

Similarly, outflow or inflow of material from/to the disk of NGC 6822
could in principle give rise to this structure.  The size and location
of the HI structure argue strongly against the outflow scenario.
Infall of material is more difficult to dismiss and could be
considered tantamount to an interaction with a small gas cloud as
mentioned above.  Finally, recent simulations show that interactions
between dark matter sub-halos and gaseous disks are unlikely creation
mechanisms for HI holes and shells \citep{kannan12}; such a scenario
in the NGC 6822 system is unlikely.

\section{Summary and Discussion}
\label{S6}

We have presented new {\it HST}/ACS imaging of six fields that span
the HI major axis of the Local Group dwarf galaxy NGC 6822.  These
positions are selected in order to probe the stellar populations
within the putative companion in the NW, the giant HI shell in the SE,
and the extended tidal material in the outer disk of the galaxy.  The
photometry reaches two magnitudes below the red clump and is extremely
sensitive to SF within the most recent 500 Myr in each of the grid
positions.

Remarkably, the SFHs of the six grid locations are essentially
identical to one another.  While the relative SFRs change from one
grid location to another, each of the six observed positions underwent
a period of heightened SF activity in the most recent $\sim$50 Myr
interval.  This interval is shorter than the dynamical timescale of
the system ($\sim$140 Myr) and argues for a galaxy-wide event at this
time.

The cumulative ancient SFHs of the six grid positions are
qualitatively similar, and reveal that NGC 6822 has formed half of its
total stellar mass within the most recent $\sim$5 Gyr.  This argues
against the inside-out growth of the disk in this system, as seen in
some other nearby galaxies. The similarity of the ancient cumulative
SFH of the putative companion and the other observed regions suggests
that this region is indistinguishable from the outer halo of NGC 6822.

This interpretation is strengthened by the lack of a coherent ancient
stellar population at the location of grid 1.  We find no over-density
of red giant branch stars at the location of the putative companion,
using either our new {\it HST} data or the previously published {\it
  Subaru} data \citep{deblok06}.  We capitalize on the photometric
depth of our {\it HST} images to derive the number of red clump stars
in each of the six grid positions; the number of red clump stars per
unit area falls off as a perfect exponential function as one moves
into the outer disk, and the number of stars in grid 1 is well within
the 1$\sigma$ error bound on this fit.  We conclude that the putative
companion is not a bona fide galaxy (i.e., one containing an ancient
stellar population).

The stellar populations within the giant HI hole show a strong and
well-populated red giant branch and red clump.  While the blue plume
is weak, there is nonetheless evidence for low-level SF activity over
the last 500 Myr.  Feedback from SF in the most recent 100 Myr is not
energetic enough to produce the HI hole. Only when we integrate over a
much longer time interval ($>$500 Myr, i.e., significantly longer than
the dynamical timescale) and assume a high feedback efficiency do we
recover sufficient feedback energy to meet the $\sim$10$^{53}$ erg
requirement derived by \citet{deblok00}.

The giant HI hole in NGC 6822 is an especially important test case for
the hypothesis that stellar feedback can create such structures.  We
find evidence for extended star formation that gives rise to this
structure.  The timescale for this feedback exceeds the dynamical time
of the galaxy.  This adds further evidence for long-lived HI
structures in solid-body rotating galaxies \citep[see,
  e.g.,][]{weisz09b,cannon11b,warren11}.  We propose that long-lived
holes and shells in the neutral ISM of dwarf galaxies are the norm;
once formed they remain coherent for long periods of time.  The lack
of dynamical processes to remove these structures allows them to
persist for hundreds of Myr in the outer disks of low-mass systems.

We consider various other creation mechanisms for the HI hole, in an
attempt to explain the enigmatic properties of the NGC 6822 system;
none is as successful as feedback from extended SF.  The overall
neutral gas morphology is indicative of a tidal interaction; the
kinematic discontinuities identified near the position of the putative
companion by \citet{deblok00} support this interpretation.  From our
{\it HST} imaging we rule out the putative companion as being a bona
fide galaxy that took part in this interaction.  However, an
interaction with a gas cloud without an ancient stellar component
(e.g., a high velocity cloud analog) remains a viable solution; the
confusion of NGC 6822 with Galactic HI may allow such a companion to
remain hidden in the current datasets.

We are observing NGC\,6822 at a unique time in its evolution.
Numerous open questions about the evolution of low-mass galaxies can
be addressed by further studies of this enigmatic object.  Of
particular value would be higher spatial and spectral resolution HI
imaging of the entire gas disk, facilitating detailed kinematic
modeling of the tidal arms, the putative companion, and the giant HI
hole.  Such data would help to differentiate between the companion
hypothesis for the NE HI cloud and a more simplistic, non-uniform HI
distribution with changing inclination angle as a function of
galactocentric radius.

\acknowledgements
 
We thank the referee, Antonio Aparicio, for insightful comments that
improved this manuscript. Partial support for this research was
provided by NASA through grant GO-12180 from the Space Telescope
Science Institute, which is operated by AURA, INC., under NASA
contract NAS5-26555.  J.M.C. and E.M.O. thank Macalester College for
research support.  This investigation has made use of the NASA/IPAC
Extragalactic Database (NED) which is operated by the Jet Propulsion
Laboratory, California Institute of Technology, under contract with
the National Aeronautics and Space Administration, and NASA's
Astrophysics Data System.

\bibliographystyle{apj}                                                 

\end{document}